\documentstyle[12pt]{article}

\newcommand{\EQ}{\begin{equation}}
\newcommand{\EN}{\end{equation}}

\renewcommand{\thefootnote}{\fnsymbol{footnote}}
\def\aprle{\buildrel < \over {_{\sim}}}
\def\aprge{\buildrel > \over {_{\sim}}}
\begin{document}
\topmargin 0pt
\oddsidemargin=-0.4truecm
\evensidemargin=-0.4truecm
\newpage
\setcounter{page}{0}
\begin{titlepage}
\begin{flushright}
Ref. SISSA 113/93/EP\\
IC/93/360\\
October 1993\\
\end{flushright}
\begin{center}
{\large NEUTRINOLESS DOUBLE BETA DECAY AND}\\
\vskip 0.5 truecm
{\large THE SOLAR NEUTRINO PROBLEM}
\end{center}
\vspace{0.2cm}
\begin{center}
{\large S.T. Petcov
\footnote{Istituto Nazionale di Fisica Nucleare, Sezione di Trieste,
 Trieste, Italy.}
\footnote{Permanent address: Institute of Nuclear Research and Nuclear
Energy, Bulgarian Academy of Sciences, BG-1784 Sofia, Bulgaria}\\}
{\em Scuola Internazionale Superiore di Studi Avanzati\\
Via Beirut 2 - 4, I-34013 Trieste, Italy} \\
\vspace{0.4cm}
{\large and}\\
\vspace{0.4cm}
{\large A.Yu. Smirnov}
\footnote{On leave from Institute for Nuclear Research, Russian Academy of
Sciences, 117312 Moscow, Russia. E-mail: smirnov@ictp.trieste.it}\\
{\em International Centre for Theoretical Physics, I-34100 Trieste, Italy}\\
\end{center}
\vspace{0.6cm}

\begin{abstract}
The MSW or vacuum oscillation
solution of the solar neutrino problem
can be reconciled with possible existence of the
$(\beta\beta)_{0\nu}$ decay with a half-life corresponding to
an effective Majorana mass of the electron
neutrino $|m_{ee}| \cong (0.1 -
1.0)~$eV. The phenomenological consequences of such a possibility
are analyzed and the implications for the mechanisms
of neutrino mass generation are considered.
\end{abstract}

\vspace{1cm}
\vspace{.5cm}
\end{titlepage}
\renewcommand{\thefootnote}{\arabic{footnote}}
\setcounter{footnote}{0}
\newpage
\section{Introduction}
\indent The results of the solar neutrino experiments
obtained so far can be considered as an indication of existence of
nonzero neutrino masses and mixing. The data are well described in terms of
neutrino resonant transitions $\nu_e \rightarrow \nu_{\mu(\tau)}$
[1], or vacuum oscillations [2]
$\nu_e \leftrightarrow \nu_{\mu (\tau)}$, with parameters
(see refs. [3] for latest analyses):
\EQ
\Delta m^2 \cong (0.6 - 1.2) \times 10^{-5}~{\rm eV}^2, ~~~
\sin^{2}2\theta \cong (0.6 - 1.4)\times10^{-2}~~~\rm{or}~~~(0.65 - 0.85),
\EN
and
\EQ
  \Delta m^2 \cong (0.5 - 1.1)\times 10^{-10}~{\rm eV}^2,
           ~~~\sin^{2}2\theta \aprge 0.75,
\EN
respectively, where $\Delta m^2 \equiv m_2^2 - m_1 ^2$,
$m_{1,2}$ being the masses of two mass eigenstate neutrinos
$\nu_{1,2}$, and $\theta$ is the lepton mixing angle in vacuum
\footnote{Let us note that solution (2) is disfavored by the data
on the neutrino burst from the supernova SN1987A [4].}.
The values of $\Delta m^2$ in (1) determine a neutrino
mass scale associated with the solar neutrino problem:
\EQ
m_{\odot} = \sqrt{\Delta m^2} \cong 3\times 10^{-3}~{\rm eV}.
\EN
\indent What are the implications of the results (1) and (2)
for the present and future searches for neutrinoless double beta
($(\beta\beta)_{0\nu}$) decay: $(A,Z) \rightarrow (A,Z+2) + e^{-} + e^{-}$
[5]? As is well known, these searches are sensitive to
the existence of massive Majorana neutrinos, $\nu_{j}$, coupled to the
electron in the weak charged lepton current.
For relatively small masses of $\nu_{j}$, $m_{j} << 30$ MeV,
the $(\beta\beta)_{0\nu}-$
decay amplitude is proportional to the $(\nu_{eL}^T\nu_{eL})$ element,
$m_{ee}$, of the neutrino Majorana mass matrix. It can be written as [6]:
\EQ
m_{ee} =~\sum_{j}^{n} \eta_{j} m_{j} |U_{ej}|^{2},
\EN
where $\eta_{j} = \frac{1}{i} \eta^{CP}_{j} = \pm 1$,
$\eta^{CP}_{j}$ is the CP-parity of the
neutrino $\nu_{j}$ \footnote{We assume for simplicity in the present study that
CP-parity is conserved in the leptonic sector and that weak interactions with
right-handed currents do not play an essential role in the
$(\beta\beta)_{0\nu}$ decay.}, and $U_{ej}$ determines the admixture of
the $\nu_{j}$ state in the $\nu_{e}$ state \footnote {$U_{ej}$ is an element
of the unitary lepton mixing matrix
defined by $\nu_{lL} = \sum_{j}^{n} U_{lj}\nu_{jL}, l = e,\mu,\tau$.}.
The results of the $(\beta\beta)_{0\nu}-$decay searches imply
the following upper bound on the value of $m_{ee}$ [7]:
$|m_{ee}| < (1 - 2)$~eV. Future experiments with enriched isotopes
$^{76}Ge$  and $^{100}Mo$ [8] will  be sensitive to values
of $|m_{ee}|$ as small as $|m_{ee}| \cong (0.1 - 0.3)$~eV.
We will refer to the effective mass having a value between the existing upper
bound and the indicated limit of sensitivity,
\EQ
m_{ee}^{obs} \cong (0.1 - 1.0)~eV,
\EN
as ``observable" effective Majorana mass.

   Simplest schemes of neutrino mass generation based on the see-saw
mechanism [9--11] do not allow to reconcile
solutions (1) and (2) of the $\nu_{\odot}$-problem with a value of $m_{ee}$
in the interval (5). Indeed, the most straightforward interpretation
of an observation of the $(\beta\beta)_{0\nu}$ decay would be that of
$\nu_e$ practically coinciding with the lightest Majorana
neutrino $\nu_1$ possessing a mass $m_1 \cong m_{ee}^{obs}$.
It follows then from the see-saw scenario that the masses of the other two
light neutrinos should lie in the intervals $m_2 \cong (1 - 100)$~ keV, and
$m_3 \cong
(0.1 - 10)$~MeV. Obviously, in this case solutions (1)
and (2) are impossible. On the other hand, e.g., solution (1) implies
$m_1 < m_2 \cong m_{\odot}$, and the see-saw mechanism with quadratic hierarchy
gives $m_3 \sim (1 - 10)$ eV in the range of
interest. However, the admixture $U_{e3}$ of $\nu_3$ in $\nu_e$
is typically predicted to be of the order of the parameter
describing the mixing between the first and the third generations in the
quark sector: $|U_{e3}| \aprle 5 \cdot 10^{-3}$ (see e.g., refs. [11]).
This gives $|m_{ee}| \aprle 2.5\times 10^{-4}$ eV $<< m_{ee}^{obs}$.

   In the present letter we analyze the phenomenological implications
of a neutrino physics solution of the $\nu_{\odot}$-problem ((1) or (2)) and
the existence of $(\beta\beta)_{0\nu}$ decay having a rate in the range of
sensitivity of the present and future $(\beta\beta)_{0\nu}-$decay
experiments [7,8]. Schemes with three light massive neutrinos are discussed in
detail. Consequences for the mechanisms of neutrino mass generation
are considered. We comment also on the possibility to accommodate two other
elements in the indicated schemes:
i) a value for the mass of one of the light neutrinos in the region of
(5 - 7) eV - such a neutrino can play the role of a "hot" dark matter
component [12], ii) oscillations $\nu_{\mu} \leftrightarrow \nu_{\tau}$
(or $\nu_{\mu} \leftrightarrow \nu_e$) with  $\Delta m^2 \cong 10^{-2} {\rm
eV}^2$  and $\sin^2 2 \theta \cong (0.4 - 0.6)$, which can explain the
suggested deficit of muon neutrinos in the atmospheric neutrino flux [13].
\section{The Case of Strong Neutrino Mass Hierarchy}

\indent Suppose that the masses of the three light Majorana neutrinos
$\nu_{1,2,3}$ obey the hierarchy relation
$m_2 << m_3$, and $m_1 << m_2$, or $m_1 < m_2$.
The $\nu_{\odot}$-problem can be solved then by $\nu_e \rightarrow\nu_{\mu}$
conversion if $m_2 \cong m_{\odot}$ and the $e-\mu$ flavour mixing corresponds
to one of the intervals in (1). A value of $|m_{ee}|$ in the interval (5)
can only be due to a sufficiently
large admixture of the $\nu_3$ state with a mass $m_3 \geq 0.1$ eV in
the $\nu_e$ state:
\EQ
|m_{ee}| \simeq  m_3|U_{e3}|^2.
\EN
\indent The values of $|U_{e3}|$ and $m_3$ are constrained
 by the null results of the oscillation experiments
performed at reactors [14,15] and accelerators [16,17].
Since $m_1 <  m_2 \cong m_{\odot}$,
the oscillation length associated with $\Delta m_{21}^2$
is much longer than the relevant source-detector distances and the
indicated oscillation experiments are sensitive only to
$\Delta m_{31}^2 = m_3^2 - m_1^2 \cong m_3^2~(\cong \Delta m_{32}^2)$.
As a consequence, the $\nu_l\leftrightarrow\nu_{l'}$ and
$\bar{\nu}_l\leftrightarrow\bar{\nu}_{l'}$ oscillation
probabilities depend only on $|U_{l3}|^2$, taking the simple form
of the probability of two-neutrino oscillations with
effective mixing parameters
$\sin^2 2\theta_{ll^{'}} = 4 |U_{l^{'}3}|^2 ~|\delta_{ll^{'}} -|U_{l3}|^2|$,
$l,l'=e,\mu,\tau$ [18].
Relation (6) can be rewritten in terms of the oscillation parameters
$\Delta m_{13}^{2} \cong m_{3}^{2}$ and $\sin^2 2\theta_{ee}\equiv
\sin^2 2\theta$. For the probability
$P(\bar{\nu}_{e}\rightarrow\bar{\nu}_{e})$ (disappearance experiments)
one has $\sin^2 2\theta = 4|U_{e3}^{2}| (1 - |U_{e3}|^{2})$, and
it follows from (6) that
\EQ
\Delta m_{13}^{2} \cong m_3^2 = \frac{4 m_{ee}^{2}}
{\left (1 - \sqrt {1 - \sin^2 2\theta}\right )^2} .
\EN
For small values of $\theta$ eq. (7) reduces to $\Delta m_{13}^{2} \approx
16 m_{ee}^2 / \sin^4 2\theta$. The parameter $\sin^2 2\theta$ determined
above enters also into the expression for the
$\nu_e \leftrightarrow \nu_{\tau}$ oscillation probability. Indeed, the
searches for $\nu_{\mu}\leftrightarrow \nu_{\tau}$ oscillations imply
$|U_{\mu 3}|^2 < 2 \times 10^{-2}$, while the negative results of the searches
for $\nu_{\mu}\leftrightarrow \nu_e$ oscillations [19] lead to the constraint
$|U_{\mu 3}|^2 < 10^{-3}/ |U_{3e}|^2$. Consequently, $|U_{\mu 3}|^2 << 1$ and
one has $\sin^2 2\theta_{e \tau} \equiv 4|U_{e3}|^2 |U_{\tau 3}|^2 =
4|U_{e3}|^2 (1 - |U_{e3}|^2 - |U_{\mu 3}|^2) \cong \sin^2 2\theta$.

 The dependence (7) of $\Delta m_{31}^{2}$ on $\sin^2 2\theta$
for different  values of $|m_{ee}|$
is shown in Fig. 1 together with the existing bounds on these two
parameters and the planned sensitivities of the future experiments [20].
As follows from Fig. 1, for $|m_{ee}| > 0.1$ eV we have $m_3 > 2$ eV and
$\sin^2 2\theta < 0.2$.
If $|m_{ee}| > 0.3$ eV one gets $\sin^2 2\theta < 0.08$ and
$m_3 \approx 4 m_{ee}/ \sin^2 2\theta > 15$ eV.
For the cosmologically interesting values of $m_3 = (3 - 7)$ eV
the allowed mixing equals $\sin^2 2\theta = (0.03 - 0.20)$.
Future oscillation experiments at accelerators (NOMAD, CHORUS [20])
will be able to cover large part of the region of interest.
In the case of negative results the allowed domains will be
around $\Delta m_{31}^2 = (3 - 5)~ {\rm eV}^2$ and
$\sin^2 2\theta = (0.1 - 0.2)$,
or at $m_3 > 30~ $eV. For $m_3 > 30~ $eV the cosmological bound implies that
$\nu_3$ should be unstable.

  A generic feature of the possibility being considered is the strong mass
hierarchy, $m_{1,2} / m_3  \aprle 10^{-3}$, and
the  relatively large mixing between the first and the third families
of leptons. This feature is a consequence of the inequality $|m_{ee}| >> m_1$
which implies a fine tuning of the elements of the $\nu_{lL}$
Majorana mass matrix, $m^{maj}$, at the level of
\EQ
\xi \equiv \frac {m_1}{|m_{ee}|} \aprle \frac {m_{\odot}}{m_{ee}^{obs}}
\aprle 10^{-2}.
\EN
Indeed, as can be shown, the ratio $\xi$ is related to the determinant
$D_m\equiv (m_{ee} m_{\tau\tau}-m_{e\tau}^2)$ of the
$\nu_{eL} - \nu_{\tau L}$ submatrix of $m^{maj}$:
$D_m /m_{e\tau}^2 \cong \xi$ \footnote{The influence of $\nu_{\mu}$
can be proven to be small.}. Consequently, $D_m \aprle 10^{-2}~m_{e\tau}^2$.
Using (6) and (8) one can write the condition for having the requisite large
mixing:
\EQ
|U_{e3}|^2  \approx \frac{1}{\xi} \frac{m_1}{m_3}
\aprge 10^2~ \frac{m_1}{m_3}.
\EN
Thus, the natural relation $|U_{e3}|^2 \aprle m_1/m_3$
is strongly violated. At the same time both the $e-\mu$ flavour mixing
(e.g., $|U_{e2}|^2 \sim 2\cdot 10^{-3}$ for the small mixing MSW solution) and
the $e-\tau$ flavour mixing are small and can
be of the order of the corresponding quark mixings.

    Consider  the implications of the above results for the
see-saw mechanism of neutrino mass generation. According to this
mechanism, the Majorana mass matrix $m^{maj}$ is given by
\EQ
m^{maj} = - m_D M_R ^{-1} m_D ^T,
\EN
where $m_D$ is the neutrino Dirac mass matrix, and
$M_R$ is
the  the Majorana mass matrix of the right-handed (RH) neutrino components
$\nu_{lR}$. We will assume that
the Dirac mass matrices of the neutrinos and the charged leptons
are similar in structure to the up and down quark mass matrices.
Correspondingly, the $e - \tau$ flavour mixing resulting only from the Dirac
neutrino and charged lepton mass matrices is exceedingly small.
An enhancement of the  $e -\tau$ mixing can take place then due to a
special structure of the Majorana mass matrix $M_R$ [21].

  It is convenient to consider the problem in the neutrino Dirac basis
$\nu_{1}^{'}, \nu_{2}^{'}, \nu_{3}^{'}$, in which the neutrino Dirac
mass matrix, $m_{D}$, is diagonal: $m_{D} = diag(m_{1D}, m_{2 D}, m_{3 D})$.
Suppose that in this basis the matrix $M_R$ has the following form
\footnote{ The Dirac basis is related to the flavour one by a
Cabibbo-Kobayashi-Maskawa type mixing matrix. This matrix generates small
$e - \mu$ and $\mu - \tau$ family mixing.}:
\EQ
M_R~=~ \left (
\begin{array}{ccc}
   M_1 & 0 & M \\
   0 & M_2 & 0\\
   M & 0 & M_3
\end{array}\right ),
\EN
where for simplicity $M$ and $M_{j}$  ($j = 1,2,3$) are considered to be
real.  The components $\nu_{2L}^{'}$ and $\nu_{2R}^{'}$ decouple from
the other neutrino components, and
$\nu_{2L}^{'} \cong \nu_{2L}\cong \nu_{\mu L}$ acquires a Majorana mass
$m_2 \cong m_{2 D}^2/M_2$. The see-saw mechanism produces an $e - \tau$
mixing
\EQ
U_{e3}~\cong~ \frac{m_{1D} m_{3D} M}
{M_{1} m_{3D}^2 - M_3 m_{1D}^2} ~ \cong ~
\frac{m_{1D}}{m_{3D}} ~\frac{M}{M_1},
\EN
and masses of $\nu_1$ and $\nu_3$
\EQ
m_1 = \frac{m_{1D}^2}{M_1}, ~~~~~~
\frac{m_{1}}{m_{3}}~=~\left(\frac{m_{1D}}{m_{3D}}\right)^2\cdot
\frac{D_M}{M_{1}^2}~
\cong~ |U_{e3}|^2 \frac{D_M}{M^2},
\EN
where $D_M \equiv M_1M_3 - M^2$ is the determinant of
the 1 - 3 submatrix of $M_R$. In eq. (12) we took into account the fact that
the rotation to the flavour basis gives a small correction to the
$e-\tau$ flavour mixing, and
have neglected $M_3m_{1D}^2$ in comparison with $M_1m_{3D}^2$
since $m_{1D}/m_{3D} \sim m_{u}/m_{t} \sim (3 - 5)\cdot 10^{-5}$,
$m_{u}$ and $m_{t}$ being the u- and t-quark masses.

 According to (12), the needed enhancement
of the $e - \tau$ mixing implies a hierarchy between the
elements of $M_R$: $M/M_{1} \aprge 10^3$. As follows from (13),
the determinant $D_{M}$ should be
small to ensure the necessary enhancement of the
mixing for small values of the ratio $m_1 /m_3$. Moreover, as can be shown
using eqs. (10) and (11),
\EQ
\frac{D_M}{M^2} = \frac{D_m}{m_{e \tau}^2} \cong \xi \aprle 3 \cdot 10^{-2},
\EN
i.e., the fine tuning of the elements of (11) is precisely the
same as in the light neutrino Majorana mass matrix $m^{maj}$. The
see-saw mechanism transforms the fine tuning problem from the
light neutrino sector to the heavy one.
This in turn implies a strong mass hierarchy of the heavy Majorana neutrino
masses:
\EQ
\frac{M_{1}^{d}}{M_{3}^{d}} = \xi \frac{M^2}{M_3 ^2}.
\EN
\indent To estimate the elements of the matrix $M_R$ we use
as an input the values $m_{1D} = (0.5 -7)$ MeV,
and $m_{3D}/m_{1D} = (0.4 - 3.0)\cdot10^4$ - typical for
charged leptons and quarks. Then from (13) - (15)
and for the maximal value $m_1 \cong m_{\odot} = 3 \cdot 10^{-3}$ eV we get
for the elements of $M_R$ (in GeV): $M_1 \sim (0.08 - 16)
\cdot 10^6$, $M \sim (0.03 - 30)\cdot 10^9 \cdot (U_{e3}/0.1)$,
$M_3 \sim (0.01 - 60) \cdot 10^{12} \cdot (U_{e3}/0.1)^2$.
With diminishing $m_1$ these masses increase as
$\propto m_1 ^{-1}$. For $m_{2D} = (0.1 - 2.0)$ GeV
and $m_2 = 3 \cdot 10^{-3}~$eV one has $M_2 = (0.25 - 200) \cdot 10^{10}~$GeV
which can be of the order of or much smaller than $M_3$.
Thus, the explanation of the inequality $m_1 << |m_{ee}|$ requires
a strong hierarchy between the elements of the matrix $M_R$,
$M_3 : M : M_1 \sim 1 : 10^{-3} :10^{-6}$. The fine tuning
of the elements of $M_{R}$ is minimal when $m_1$ is comparable with $m_2$;
the spectrum in this case has a rather peculiar form: $m_3 >> m_2 \sim m_1$.
Let us note that the higher order corrections to the ratio
$D_M/M^2$ affecting condition (14) are estimated to be $\sim
10^{-5}$.

  We shall outline next several approaches permitting to explain the required
properties of the matrix $M_R$. Its texture, hierarchy of elements as well as
the smallness of the determinant of the 1 - 3 submatrix can be a consequence
of a certain family symmetry $G$. The elements of the matrix $M_R$ can appear
as bare mass terms conserving $G$, or/and can be generated by
couplings to new scalar fields, $\sigma$, singlets of $SU(2)_L\times U(1)$.
These scalar fields have, in general, nonzero $G$-charges, $G_{\sigma}$, and
acquire nonzero vacuum expectation values, $\sigma_{0}$,
thus spontaneously breaking the symmetry $G$. Let us consider several
simplest  possibilities.

1. The texture (11) can be easily generated by bare mass terms and by
interactions with a field $\sigma$ if, e.g., $G = U(1)$,
the $G$-charges of the neutrinos are
$G(\nu_{1L,R}^{'},\nu_{2L,R}^{'},\nu_{3L,R}^{'}) = (1, 0, -1)$,
$G_{\sigma} = 2$, and $G(\Phi) = 0$, where $\Phi$ is the standard
Higgs doublet. However, condition (14) implies a fine tuning between the bare
mass M and the contributions from the interaction with $\sigma$.

 Instead of bare mass terms for $\nu_{jR}^{'}$ one can introduce couplings
with additional scalar fields.
Suppose that: $G = SU(2)$, the scalar fields $\sigma_{11}, \sigma_{13},
\sigma_{33}$ form a triplet {\bf $\sigma$} of $G$, $\nu_{1R}^{'}$ and
$\nu_{3R}^{'}$ are in a doublet of $G$ and interact with {\bf $\sigma$},
whereas $\nu_{2R}^{'}$ is a $G$-singlet. The Higgs potential can be easily
arranged in such a way that only one component of the triplet ${\bf \sigma}$
acquires a nonzero vacuum expectation value. If, e.g., this
component coincides with $\sigma_{33}$, only the $M_{33}$ element
of the 1 - 3 submatrix of $M_R$ will be nonzero. In general, the basis in
which only one component of the triplet develops a nonzero vacuum expectation
value differs from the neutrino Dirac basis
by a rotation on a certain angle $\alpha$. Thus, the vacuum expectation values
of the components of the triplet in the neutrino Dirac basis will be related
as $\sigma_{33}$ : $\sigma_{13}$ : $\sigma_{11}$ =
$\cos^2 \alpha$ : $\cos \alpha \sin \alpha$ : $\sin^2 \alpha$.
Correspondingly, the same relation will take place
for the components of the 1 - 3 submatrix of $M_R$. For
$\sin \alpha \cong 10^{-3}$, one reproduces the necessary hierarchy of the
elements and, moreover, the determinant $D_M$ is identically zero.
A sufficiently small value of $D_M$ can then arise as a result of weak
violation of the symmetry $G$.

   2. Suppose that the symmetry $G = U(1)$ is broken by the vacuum expectation
value of {\it only one} singlet scalar field $\sigma$ which carries a charge
$G_{\sigma}$. The value of the $M_i$ element of the matrix $M_R$ (11) can be
related to the $G$-charge $G_i$  of the corresponding mass term operator as
follows
\EQ
M_i = M_0 \left( \frac{h\sigma _0}{M_0} \right) ^{G_i/G_{\sigma}},
\EN
where  $M_0$ is a (bare) mass parameter and $h$ is the constant of the
$\nu_{jR}^{'}$ and $\sigma$ Yukawa coupling. Prescribing
$G(\nu_{1L,R}^{'},\nu_{2L,R}^{'},\nu_{3L,R}^{'})~=~ (1, 1/2, 0)$
and $G_{\sigma} = 1$, we get: $M_3 = M_0$, $M = M_2 = M_0
\left(\frac{h\sigma_0}{M_0}\right)$,
$M_1 = M_0 \left(\frac{h\sigma_0}{M_0}\right)^2$.
All other elements of $M_R$ are zero since they have no $G$-invariant
interactions with $\sigma$. The ansatz (16) ensures the needed relation
between the elements of $M_{R}$ ($D_M = 0$). For $\sigma_0/M_0 \cong
10^{-3}$  we get the requisite hierarchy of the elements of $M_R$.
Weak violation of the $G$ symmetry at the scale of ($10^3 - 10^4$) GeV will
generate the mass of the lightest RH component.
The models reproducing the ansatz (16) should
contain, e.g., an additional scalar field $\sigma^{'}$ with
$G_{\sigma^{'}} = 2$, having specific mass
and couplings (e.g., $M_{\sigma^{'}} = M_0$,  trilinear coupling
$hM_0 \sigma\sigma \sigma^{'}$, and Yukawa coupling to $\nu_{1R}$ with
a constant $h^{'} = h$).

   3. One can make use of the fact that the 1 - 3 submatrix of $M_R$
is diagonalized by a rotation on angle $\theta \sim 10^{-3}$ and its
eigenvalues obey the hierarchy $M_1^{d}/M_3^{d} \cong 10^{-9}$.
Suppose that due to a certain symmetry $G$ the matrix $M_R$ is
diagonal and $M_2^d \aprle M_3^d$, whereas $M_1^d = 0$. (The RH component
$\nu_{1R}$ acquires a mass due to weak violation of $G$, so that
$M_1^d << M_2^d$). The Yukawa interactions
of the standard Higgs doublet field can violate $G$, so that the neutrino
Dirac basis differs from the basis in which $M_R$ is diagonal by
a transformation $S$. It can be shown that the
needed enhancement of the $e - \tau$ flavour mixing is achieved if
$S$ has a structure similar to that of the Cabibbo-Kobayashi-Maskawa matrix.
\section{Three Mass-Degenerate Neutrinos}

\indent  Consider a highly degenerate neutrino mass spectrum:
$m_1 \simeq m_2 \simeq m_3 = m_0$, with $m_0 \aprge 0.1$ eV. The effective
neutrino mass parameter can be written then as
\EQ
m_{ee} = m_0 \sum_{i = 1,2,3} |U_{ei}|^2 \eta _i ,
\EN
and, in general, all three components give appreciable
contributions in the sum. The solar neutrino deficit is explained by
$\nu_e$ conversion to, e.g., $\nu_{\mu}$ if
$|m_2 - m_1|  = \Delta m_{21}^2 / 2m_0 \cong m_{\odot}^2/ 2m_0
\aprle (0.5 - 5)\cdot 10^{-5}~$eV, or by vacuum oscillations provided
$|m_2 - m_1| \cong (1 - 5) \cdot 10^{-10}$ eV. It is
possible also to describe the
atmospheric neutrino deficit in this case as caused by the oscillations
$\nu_{\mu} \leftrightarrow \nu_{\tau}$, which would require
$(m_3 - m_2) \cong 10^{-2} {\rm eV}^2 /2 m_0 = (0.5 - 5)\cdot 10^{-2}$ eV
\footnote{Such a possibility has been considered in another context in
[22].} .
With such a highly degenerate neutrino mass spectrum the reactor and
accelerator data [14-17] do not impose any limits on the mixing.

  There are several phenomenological possibilities depending on the relative
signs of $\eta_j$, $j=1,2,3,$
and on the values of the mixing parameters. If $\nu_1$, $\nu_2$, and $\nu_3$
have the same CP-parity, one obtains
taking into account the unitarity of the lepton mixing matrix: $m_{ee} = m_0$.
Consequently, the existing data on the $(\beta\beta)_{0\nu}$ decay implies:
$m_0 < (1 - 2)$ eV. If $\eta_j$ have different signs,
mutual compensation of the contributions in $m_{ee}$ will take
place. The compensation is substantial for large mixing, and as a result of it,
$m_0$ can be considerably larger than $|m_{ee}|$.

  Consider first the case when the admixture of $\nu_3$ in $\nu_e$ is small:
$|U_{e3}|^2 << 1$. Using the unitarity condition one can write the
effective neutrino mass parameter as
\EQ
|m_{ee}| \approx m_0 \left|~1 + (\eta_1 \eta_2 - 1) |U_{e1}|^2~\right|.
\EN
If $\eta_1 = \eta_2$, we have
$m_0 = |m_{ee}|$ and the bounds on $m_0$ are independent of the mixing
(Fig. 2): they are the same for all solutions of the $\nu_{\odot}$-problem.
For $\eta_1 \eta_2 = -1$ the compensation takes place; it becomes stronger as
$|U_{e1}|^2$ approaches 1/2 (see (18)). Correspondingly, when
$|U_{e1}|^2 \rightarrow  1/2$ the
bounds on $m_0$ (both the upper bound from existing data on the
$(\beta\beta)_{0\nu}$ decay, and the lower bound following from the
observability condition) increase.
For the large mixing solution (1), the maximal allowed value of
$\sin^2 2\theta$ is (0.85 - 0.90); it gives
$|m_{ee}| = (0.4 - 0.5) m_0$ and a maximal allowed value of
$m_0 \approx (4 - 5)$ eV. For the vacuum oscillation solution
values of $\sin^2 2\theta$ up to the maximal one are not excluded,
making possible even stronger cancellation and values of
$m_0$ up to the upper bound on the electron neutrino mass obtained in the
tritium $\beta$-decay experiments: $m_0 \aprle  7$ eV [23]. In the case
of the small mixing solution (1) one has $m_0 \cong |m_{ee}|$
independently of the relative CP parities of the neutrinos $\nu_j$.

   Let us consider now the possibility of the third neutrino $\nu_3$
having an appreciable
admixture in $\nu_e$. The resonant conversion of the solar neutrinos will be
effective if the lightest  component $\nu_1$ dominates in $\nu_e$, i.e., if
$|U_{e1}| > |U_{e2}|, |U_{e3}|$. Therefore the maximal compensation effect
takes place when $\eta_2 = \eta_3 = - \eta_1$. The expression for $m_{ee}$
reduces to (18). For an appreciable admixture of the $\nu_3$ state in
$\nu_e$, the region of the large mixing angle solution extends to
$\sin^2 2\theta \cong 0.95$ [24,25].
The allowed region of parameters enlarges (see Fig. 2) and values
of $m_0$ up to the experimental upper limit of 7 eV are possible.

  We shall analyze next the relevant implications for the mechanisms of
neutrino mass generation. If the CP-parities of all Majorana neutrinos are
the same ($\eta_1 = \eta_2 = \eta_3$), the Majorana mass matrix
$m^{maj}$ in the flavour basis can be represented as
\EQ
m^{maj}~=~m_0\cdot\hat{I} + \delta \hat{m},
\EN
where $\hat{I}$ is the unit matrix and $\delta\hat{m}$ is a "small"
correction matrix generating the neutrino mass splitting and the mixing.
The neutrino conversion (oscillation) parameters are determined then
by the matrix
\EQ
H~\cong~\frac{m_0}{E}\cdot\delta\hat{m}.
\EN
In order to have $\Delta m^2_{21},~{\rm or/and} ~\Delta m^2_{31} \aprle
10^{-5}~{\rm eV}^2$,
the elements of $\delta\hat{m}$ (at least some of them) should be of the
order of or smaller than $m_{\odot}^2/m_0 \sim (10^{-4} - 10^{-5})$~eV.
It is natural to suggest that $m_0$, the main contribution in $m^{maj}$,
is generated by interactions which preserve a family symmetry $G$,
whereas $\delta \hat{m}$ results
from $G$-breaking interactions. Suppose that the lepton doublets $\psi_{lL}$
form a triplet of a horizontal group $G$ (e.g., $G = SO(3)$). The
$m_0$ - term in (19) can be generated by $G$ invariant interaction of
the Higgs doublets with a superheavy $SU(2)_{L}$ triplet Higgs field
$\Delta_{ij}$  $(\Delta^0_{11},\Delta^{-}_{12},\Delta^{--}_{22})$ carrying a
zero $G$-charge,
\EQ
h_0\sum_{l = e, \mu, \tau}
\psi^{T}_{lL} C \psi_{lL}\Delta^{\dagger} + h.c., \EN
where $h_0$ is a Yukawa constant. A sufficiently small vacuum
expectation value $\Delta_0$ of $\Delta$ can be induced by a quartic
coupling of $\Delta$ with the standard Higgs doublet $\Phi$ and an additional
singlet field $\sigma$: $\Delta\Phi\Phi\sigma$. The singlet $\sigma$ can
naturally acquire a vacuum expectation value $\sigma_0 >> \Phi_{0}$,
$\Phi_{0}$ being the vacuum expectation value of $\Phi$, and one obtains:
$\Delta_0 \sim (\Phi_{0})^{2}/\sigma_0$ [26]. Interaction (21) generates then
the mass $m_0 = h_0 \Delta_0$. For $\sigma_0
\sim M \cong (10^{12} - 10^{14})~$GeV
and $h_0 \cong (10^{-2} - 10^{-1})$ one obtains $\Delta_0~\cong
(1 - 10^{2})~$eV and $m_0 = (0.1 - 1.0)$ eV.

  The contribution to the ``small" term in (19), $\delta
\hat{m}$, arises from one loop correction induced by the Yukawa couplings of
leptons with $\Phi$. One finds that, e.g.,
$\delta\hat{m}_{\tau\tau} \sim m_0 \frac{G_{F}m_{\tau}^2}
{8 \sqrt 2 \pi^2}
\ln \frac{M_{\Delta}^2}{M_{W}^2} \aprle 3 \cdot 10^{-5} m_0$. The matrix
$\delta\hat{m}$ may be generated, e.g., by Planck-scale interactions [27]:
\EQ
\delta\hat{m}_{ll'}~=~\frac{\alpha_{ll'}}{M_{Pl}}~
   \psi^{T}_{lL} C \tau_2 \bar{\tau}\psi_{l'L}\Phi^{T} \tau_2
      \bar{\tau}\Phi + h.c.,
{}~~l,l'= e,\mu,\tau,
\EN
where $M_{Pl}$ is the Planck mass. For $\alpha_{ll'}\sim 1$, one has
$\delta\hat{m}_{ll'}\sim 10^{-5}$ eV, which gives $\Delta m^2_{21}
\sim m_{\odot}^2$. The constants $\alpha_{ll'}$
can be flavour universal [27]. One
obtains in this case the large mixing angle MSW solution (1). If $\alpha_{ll'}$
have values spread over an order of magnitude, $\alpha_{ll'}\sim (10^{-1} -
1)$, the small mixing angle solution is reproduced.

  The matrix $\delta\hat{m}$ can appear also as a result of the see-saw
mechanism. The relevant parameters
can take values permitting to explain both the solar neutrino problem
and the atmospheric neutrino deficit. For $M_R~=~M_0\cdot\hat{I}$,
with $M_0 = 10^{11}$ GeV one gets $\delta m_{\mu \mu} \sim 10^{-5}$ eV,
and $\delta m_{\tau\tau} \sim 10^{-2}$ eV, so that
$\Delta m_{21}^2 \sim 2 m_0 \delta m_{\mu \mu}$
and $\Delta m_{31}^2 \sim 2 m_0 \delta m_{\tau \tau}$
are precisely in the requisite intervals.

   In the schemes considered above neutrinos acquire masses due to couplings to
superheavy particles at large energy scales, respecting
a family symmetry
(interactions with $\Delta$,
Majorana mass terms of the RH neutrino components, etc.),
whereas the masses of the charged particles
as well as the lepton mixing are generated by the
interactions with the standard Higgs doublet, which break this
symmetry. An alternative possibility is that the Higgs doublet couplings also
respect the family symmetry and neutrinos acquire equal masses via the
see-saw mechanism. The hierarchical structure
of the mass spectrum of the charged particles (quarks, charged leptons)
may result from a certain pattern of the vacuum expectation values of
the scalar horizontal multiplets.
Such a possibility can be realized in terms of the universal see-saw
mechanism [28].
\section{Two Mass-Degenerate Neutrinos}

\indent Let us discuss the modification of the above scenario with only two
almost degenerate neutrinos,
say $\nu_1$ and $\nu_2$, having masses in the
region of interest, $m_1 \cong m_2 \cong m_0 \aprge 0.1$ eV.
The third neutrino $\nu_3$ has a mass which can differ considerably from $m_0$.
As before, the $\nu_{\odot}$-deficit
is explained by $\nu_e \ - \nu_{\mu}$ conversion (or oscillations) induced
by a small mass difference of the degenerate neutrinos: $(m_2 - m_1)/m_0\cong
m_{\odot}^2/(2m_0^2)\aprle 10^{-3}$. The effective Majorana neutrino mass
parameter can be written as
\EQ
m_{ee} = m_{12} + \eta_3 m_3 |U_{e3}|^2,
\EN
where $m_{12} \equiv \eta_2 m_0 \left [ ~1 + (\eta_1 \eta_2 - 1)
|U_{e1}|^2 \right]$
is the contribution due to the degenerate neutrinos. If the CP-parities of
$\nu_1$ and $\nu_2$ are opposite, i.e., if $\nu_1$ and $\nu_2$ form a
pseudo-Dirac neutrino [29,30], then
$|m_{12}| = m_0 \left | 1 - 2 |U_{e1}|^2 \right |$.
Depending on the value of $m_3$ we get different phenomenological implications.

   For $m_3 << m_0$ one has $\Delta m^2_{13} \sim \Delta m^2_{23}
\sim m^2_0 \aprge 10^{-2} {\rm eV}^2 $ and the predicted
$\nu_e \leftrightarrow \nu_{\tau}$ and
$\nu_{\mu} \leftrightarrow \nu_{\tau}$ oscillation effects
can be in the range of sensitivity of the future experiments [20]. The
existing limits from the oscillation searches
imply an upper bound on $|U_{e3}|$ which makes the contribution
due to $\nu_{3}$ in $m_{ee}$ negligibly small. Thus, in what regards
the $(\beta\beta)_{0\nu}$ decay, the present case reduces to
the two neutrino case (18) analyzed in the previous section.

    If $m_3 \aprge m_0$, there are several possibilities.
The case $|m_{12}| < 0.1$ eV is equivalent, as far as the
$(\beta\beta)_{0\nu}$ decay is concerned, to the one considered
in section 2. New features appear when the contributions in
$m_{ee}$ due to $\nu_3$ and the $m_{12}$ term are
comparable. For values of $m_3$ below the cosmological bound,
$m_3 \aprle 30$ eV, one has, taking into account the results of the
oscillation experiments [14--17], $m_3 |U_{e3}|^2 \aprle 0.5$ eV. The
allowed region shown in Fig. 2 is shifted to smaller (larger) values of $m_0$
when $m_{12}$ and $\eta_3 m_3$ have the same sign (opposite signs).
In particular, $|m_{12}|$ can be
close to the upper bound on the electron antineutrino mass: $|m_{12}|\approx
(5 - 7)$ eV. In the latter case the term $m_{12}$ in $m_{ee}$ should be
cancelled by the
contribution from $\nu_3$, which implies $m_3 |U_{e3}|^2 \approx (4 - 8)$ eV.
Using the bound on $|U_{e3}|^2$ from the
oscillation experiments one finds $m_3 \aprge (200 - 400)$ eV.
Obviously, the neutrino $\nu_3$ should be unstable.

 Finally, the case when $m_3 \aprge |m_{ee}|$, $m_0 << |m_{ee}|$ is
equivalent from a phenomenological point of view to the one discussed in
section 2.

 Two highly degenerated neutrinos  appear in a theory
with a horizontal symmetry $G$, with two families
having nonzero $G$-charges and the third family being a singlet of $G$.
This possibility can be realized if neutrinos acquire masses due to an
interaction with the triplet {\bf $\Delta$} as in eq. (21) (now two families
form a doublet of $G$), or in terms of the see-saw mechanism.
Suppose $G = U(1)$ and the neutrinos have
$G$-charges $(\frac{1}{2}, -\frac{1}{2}, 0)$, then $G$-conserving
bare mass terms and/or interactions with scalar singlets having
zero $G$ charges will generate the matrix $M_R$ with only nonzero
elements $M_{12} = M_{21}$ and $M_{33}$. Such a matrix
leads via the see-saw mechanism to a light Majorana
neutrino with mass $m_3$ and a light pseudo-Dirac neutrino with
mass $m_0$, given by
\EQ
m_3 = \frac{m_{3D}^2}{M_{33}}, ~~~
m_0 = \frac{m_{1D} m_{3D}}{M_{12}}.
\EN
For $M_{33} = 10 M_{12}$ we obtain from (24) $m_3 \approx 10$ eV and
$m_0 \sim (0.01  - 0.1)$ eV (and even a stronger hierarchy is possible).
Small mass splitting between the two Majorana components of the
pseudo-Dirac neutrino is generated by weak violation of the $G$-symmetry
by  $M_R$.
If the basis in which the neutrino states have definite
$G$-charges differs from the flavour basis by a rotation on an
angle $\theta_b$ of the $\nu_1$ and $\nu_2$ states, one finds that the mixing
angle appearing
in the expressions for $\nu_e$ and $\nu_{\mu}$ states is given by
 $\tan 2\theta \cong 1/ \tan 2\theta_b$. For the
value $\theta_b \sim\theta_c \cong 13^0$ we get
$\sin ^2 2\theta \cong 0.8$, i.e., such a scheme can
accommodate the large mixing solution of the $\nu_{\odot}-$problem.

 Two almost mass-degenerate neutrinos may appear in the models with radiative
mechanism of neutrino mass generation, like the Zee model [31].
Simplest models predict typically near to maximal mixing between
the mass-degenerate states and maximal mixing of two flavour neutrinos. This
is disfavored by the MSW solution of the $\nu_{\odot}$ problem.
Moreover, the two almost mass-degenerate Majorana neutrinos usually have
opposite CP parities, their contribution to $m_{ee}$ has the form
$m_{12} \sim \Delta m^2 / 2m_0$ and turns out to be small.
One can avoid the maximal mixing and the suppression of
$m_{ee}$ by modifying the Zee model, e.g., by introducing Higgs
doublets with couplings to the lepton fields which are antisymmetric in
the flavour indices.

\section{Conclusions}

  In the present paper we have limited our discussion to the case of only
three light massive Majorana neutrinos. One could consider also schemes with
larger number of light neutrino states, for instance,
with light sterile neutrinos $\nu_{sL}$. This, obviously, opens up new
possibilities for reconciliation of an observable
$(\beta\beta)_{0\nu}$ decay rate with the solar and atmospheric neutrino
data.

   A $(\beta\beta)_{0\nu}$ decay with half-life in the range of the sensitivity
of the future experiments can be caused by a mechanism
not directly related to the exchange of light Majorana neutrinos,
for example, by weak interactions with right-handed
currents, exchange of heavy neutrinos ($m > 30~$MeV) or other heavy
neutral Majorana particle(s),  etc.
[5]. In these cases the usual schemes of neutrino mass
generation need not be modified in order to obtain the MSW or the vacuum
oscillation solution of the $\nu_{\odot}-$problem.

  In conclusion, we have shown that from a phenomenological
point of view it is not difficult to reconcile the  particle
physics solution of the $\nu_{\odot}-$problem and an observable
$(\beta\beta)_{0\nu}$ decay induced by exchange of Majorana neutrinos with
an effective neutrino mass $(0.1 - 1.0)$ eV. Moreover, some proposed schemes
can accommodate also the solution of the atmospheric neutrino problem
and contain at least one neutrino state with mass in the
cosmologically interesting range. Neutrino oscillations
which can be probed in future laboratory
experiments are predicted in schemes with strong hierarchy between the neutrino
masses. However, as we show, such a reconciliation will have
serious implications for the mechanisms of neutrino mass generation.
The implications depend on the relation between the
mass of the lightest neutrino $m_1$,  the solar $\nu_{\odot}$-problem mass
scale $m_{\odot}$, and the effective Majorana mass $m_{ee}$.
There are two extreme cases.

  If $m_1 < m_{\odot} << |m_{ee}|$ {\it (strong mass hierarchy)}, the
elements of the LH flavour neutrino Majorana mass matrix should be "tuned"
to obey a certain relation
with a relative precision of $m_1/|m_{ee}|\aprle 10^{-2}$. In the see-saw
mechanism, such a
relation may result from a similar one between the elements
of the mass matrix of the RH neutrinos. The later could be
explained in turn by the presence of a family symmetry and strong
hierarchy of masses of the RH neutrinos.

  The case $m_1 \aprge |m_{ee}|$ implies a {\it strongly degenerate} light
neutrino mass spectrum. Such a spectrum can be produced
by new interactions at high mass scale which respect a certain
family symmetry. The latter can be weakly broken by low mass scale
interactions of the usual Higgs doublet, or by gravitational
effects. In the intermediate case $m_1 < m_{\odot} < |m_{ee}|$ both
the tuning condition should be satisfied and two light
neutrinos should be almost mass-degenerate. The stronger the
degeneracy the weaker the "tuning", and vice versa.

   Part of the work of S.T.P. for the present study was done during a visit
to the L.P.T.H.E. and L.A.L. of the Centre d'Orsay in the spring of 1993.
He wishes to thank L. Oliver and S. Jullian for stimulating discussions, and
O. P\'ene and the other members of L.P.T.H.E. for the kind hospitality extended
to him during his visit. A.Yu.S. would like to thank Z. Berezhiani,
G. Dvali and G. Senjanovic for illuminating discussions.
The work of S.T.P. was supported in part by
the Bulgarian National Science Foundation via grant PH-16.

\newpage
\centerline{\bf Figure Caption}
\vspace{0.8cm}
\begin{description}
\item{Fig. 1.} The dependence of $\Delta m^2 \cong m_{3}^{2}$ on
$\sin^{2}2\theta~=~4|U_{e3}|^2~(1 - |U_{e3}|^2)$
(solid lines) for different values of $|m_{ee}|$ (see eq. (7)). The
regions of parameters excluded by the G\"{o}sgen [14], BEBC [16] and IPR [15]
oscillation experiments are depicted. Shown are also the regions of sensitivity
of the future oscillation experiments CHORUS, NOMAD, P-803 and P-860 [20].
\item{Fig. 2.} Values of the mass $m_0$ of the degenerate neutrinos
and the mixing parameter $|U_{e1}|^2$ for which the MSW and vacuum oscillation
solutions of the $\nu_{\odot}-$problem can be reconciled with observable
Majorana mass $|m_{ee}| = (0.1 - 1.4)$ eV.  Solid lines correspond to two
neutrino contributions in $m_{ee}$ and to two-neutrino
oscillations/conversions.
The regions of the large mixing MSW
solution are hatched; the small mixing solution is shown as a vertical
line at $|U_{e1}|^2 \cdot \eta_1 \cdot \eta_2 \cong \pm 1$.
For an  appreciable contribution of the third neutrino state in $m_{ee}$ the
regions are larger: the dashed lines correspond to the case of three
degenerate neutrinos, the dotted lines correspond to the case
of large $m_3$, so that $m_3~|U_{e3}|^2 = 0.5~$eV. The upper bound on
the electron antineutrino mass from the tritium experiments is also shown.
\end{description}
\end{document}